\documentstyle[12pt,epsf,epsfig,psfig]{article}
\def\x{\chi}

\def\cc{\langle{\bar \psi} \psi \rangle}

\def\be{\begin{equation}}
\def\ee{\end{equation}}

\def\lsim{\raise0.3ex\hbox{$<$\kern-0.75em\raise-1.1ex\hbox{$\sim$}}}
\def\gsim{\raise0.3ex\hbox{$>$\kern-0.75em\raise-1.1ex\hbox{$\sim$}}}

\def\NP{{ Nucl.\ Phys.\ }}
\def\PL{{ Phys.\ Lett.\ }}
\def\PR{{ Phys.\ Rev.\ }}

\def\PRL{{ Phys.\ Rev.\ Lett.\ }}

\begin{document}

\noindent July 17, 2000~ \hfill BI-TP 2000/25

\vskip 1.5 cm

\centerline{\Large{\bf Deconfinement through Chiral Symmetry}}

\bigskip

\centerline{\Large{\bf Restoration in Two-Flavour QCD}}

\vskip 1.0cm

\centerline{\bf S.\ Digal, E.\ Laermann and H.\ Satz}

\bigskip

\centerline{Fakult\"at f\"ur Physik, Universit\"at Bielefeld}
\par
\centerline{D-33501 Bielefeld, Germany}

\vskip 1.0cm

\noindent

\centerline{\bf Abstract:}

\medskip

We propose that in QCD with dynamical quarks, colour deconfinement
occurs when an external field induced by the chiral
condensate strongly aligns the Polyakov loop. This effect sets in at the
chiral symmetry restoration temperature $T_{\chi}$ and thus makes
deconfinement and chiral symmetry restoration coincide. The predicted
singular behaviour of Polyakov loop susceptibilities at $T_{\chi}$ is
shown to be supported by finite temperature lattice calculations.

\vskip 1.0cm

\bigskip

Finite temperature QCD at vanishing overall baryon density leads to two
well-defined phase transitions. In the limit of infinite bare quark
mass, for $m_q \to \infty$, one obtains pure $SU(N)$ gauge theory; here
deconfinement occurs when the global center $Z_N \in SU(N)$ symmetry of
the Lagrangian is spontaneously broken. The expectation value $L(T)$ of
the Polyakov loop constitutes the order parameter for this transition,
analogous to the magnetization $m(T)$ in $Z_N$ spin theories. There
exists a critical temperature $T_d$, with $L(T) = 0 ~~\forall~~T \leq
T_d$ and $L(T) > 0~~\forall~~T>T_d$. The introduction of dynamical
quarks ($m_q < \infty$) breaks this symmetry explicitly, and now $L(T) >
0~~\forall~~T > 0$.

\medskip

For $m_q=0$, the chiral symmetry of the Lagrangian is spontaneously
broken at low temperatures and restored for $T\geq T_{\x}$. The
chiral condensate $K(T)\equiv \cc$ is the order parameter for this
transition, with $K(T) > 0~~\forall~~T<T_{\x}$ and $K(T)=0~~\forall~~
T\geq T_{\x}$. A non-vanishing quark mass breaks chiral symmetry
explicitly, and for large $m_q$, the temperature variation of the
chiral condensate becomes completely smooth.

\medskip

One might consider the inverse bare quark mass to play such a symmetry
breaking role for deconfinement, which would imply that for $m_q \to 0$, 
the temperature variation of the Polyakov loop expectation value would 
also become
smooth \cite{HKS}. In lattice studies it is found, however, that this
is not the case \cite{KL}: even for $m_q \to 0$, the Polyakov loop
varies sharply with temperature and the corresponding susceptibility 
($\langle L^2 \rangle - \langle L \rangle^2$) peaks sharply at the chiral 
restoration temperature $T_{\x}$, which is considerably lower than $T_d$. 
In somewhat loose terminology, one describes this situation by saying that
in QCD for $m_q \to 0$, deconfinement and chiral symmetry restoration
coincide. The aim of this paper is to elucidate the underlying reasons
for such behaviour and to show that chiral symmetry restoration in fact
drives the observed Polyakov loop variation; in particular, it leads to
singular behaviour for specific Polyakov susceptibilities.

\medskip

Conceptually, our starting point is the idea that it is the inverse
of the constituent quark mass $m_Q$, rather than the bare mass $m_q$,
which acts as an external field for the $Z_N$ symmetry \cite{Gavai}.
Evidently there does not exist a clear definition of $m_Q$; for the
present discussion, we take it to be determined by the (non-Goldstone)
hadron masses. For our actual conclusions, however, we shall avoid the
intermediate step of constituent quarks altogether. In the limit
$m_q \to 0$, $m_Q$ remains finite in the temperature region in which
chiral symmetry is spontaneously broken, since the hadron masses do. If
the hadron masses are related to the chiral condensate, e.g., if the
nucleon mass is given by \cite{Ioffe}
\be
M_n \sim 3~m_Q \sim \cc^{1/3}, \label{1}
\ee
then chiral symmetry restoration with $\cc \to 0$  will lead to a
sudden increase of the external field $H \sim 1/m_Q$, forcing a large
explicit breaking of $Z_N$ and hence a corresponding alignment of the
Polyakov loop. We want to argue that it is this effect which causes a
sharp variation of the Polyakov loop at $T_{\x} < T_d$, with the
functional form of the variation determined by chiral symmetry
restoration, which is in general different from that obtained in pure
gauge theory at $T_d$.

\medskip

We begin by summarizing the main features of the critical behaviour for
the two transitions, based on QCD with two species of dynamical quarks
and colour $SU(3)$. The order parameter for chiral symmetry restoration,
$K \equiv \cc$, vanishes for $m_q=0$ as $T \to T_{\x}$ from below,
\be
K(T,m_q=0) \sim (T_{\x}-T)^{\beta},~~T \lsim T_{\x} \label{2}
\ee
while at $T=T_{\x}$, it vanishes as
\be
K(T_{\x}=0,m_q) \sim m_q^{1/\delta} \label{3}
\ee
when $m_q \to 0$. The chiral quark mass susceptibility
\be
\x_m^K(T,m_q) = \left( {\partial K \over \partial m_q}
\right)_T \sim ~ \langle  (\psi{\bar \psi})^2 \rangle - \langle
\psi{\bar \psi} \rangle^2~ \sim~|T-T_{\x}|^{-\gamma}
\label{4}
\ee
diverges at $T = T_{\x}$ for $m_q \to 0$. The critical behaviour at the
chiral symmetry restoration point is thus specified in terms of
critical exponents $\beta,~\delta,~\gamma,~...$. For two massless quark
flavours, the transition is conjectured to belong to the $O(4)$
universality class \cite{PW}, which would determine the exponents.
Present lattice calculations cannot yet determine if this conjecture is
correct \cite{Aoki,EL}.

\medskip

In the limit $m_q \to \infty$, we recover pure $SU(3)$ gauge theory, for
which the deconfinement transition is of first order, as it is for the
corresponding $Z_3$ spin theory \cite{SY}. The resulting discontinuity
at $T=T_d$ persists in $L(T,m_q)$ for sufficiently large $m_q$; but
below some quark mass value $0< m_q^d < \infty$, $L(T,m_q)$ and all its
derivatives presumably vary smoothly with temperature. We want to argue,
however, that for $m_q \to 0$, certain derivatives of $L(T,m_q)$ will
become singular.

\medskip

As indicated above, we now take the $Z_N$ structure of QCD to be that
of $Z_N$ spin theory in an external field $H$, determined by the
constituent quark mass. From Eq.\ (\ref{1}), $H$ and hence also the
Polyakov loop thus become functions of $K$: the quark mass dependence of
$L$ enters through $K(T,m_q)$, with $L(T,K)$. In spin systems
with external field, the magnetization $m(T,H)$ is for $H\not=0$ an
analytic function of $T$ and $H$. We therefore assume that $L(T,K)$ is
for $\infty>K>0$ an analytic function of $T$ and $K$, with the quark mass
dependence of $L$ entering through $K(T,m_q)$.

\medskip

As a consequence, the critical behaviour of the chiral condensate will
be reflected in the behaviour of the Polyakov loop at $T=T_{\x}$. A first 
hint that this is the case is seen in \cite{Rajan}-\cite{MTc}. For 
four flavours
of light quarks, the chiral transition becomes first order, with a
discontinuity in $K(T,m)$ at $T_{\x}$ for $m_q < m_q^{\x}$ smaller than
some `endpoint value' $m_q^{\x}$. The corresponding Polyakov loop $L(T)$
also shows a discontinuity at this temperature, as expected from our
considerations. Note that here we have a first order transition at
$T=T_{\x}$ for $0 \leq m_q < m_q^{\x}$, induced by chiral symmetry
restoration, and another first order transition at $T=T_d > T_{\x}$ for
$m_q > m_q^d$, induced by spontaneous $Z_N$ symmetry breaking. For
$m_q^{\x} < m_q < m_q^d$, $L(T,K)$ varies smoothly with $T$.

\medskip

In the case of two dynamical quark flavours, the chiral transition is
presumably continuous. From the $L$-variation
\be
dL = \left( {\partial L \over \partial T} \right)_K dT +
\left( {\partial L \over \partial K} \right)_T dK
\label{5}
\ee
we then find that the Polyakov loop susceptibilities
\be
\x^L_m = \left( {\partial L \over \partial m_q} \right)_T =
\left( {\partial L \over \partial K} \right)_T
\left({\partial K \over \partial m_q} \right)_T
\label{6a}
\ee
and
\be
\x^L_T = \left( {\partial L \over \partial T} \right)_{m_q} =
\left( {\partial L \over \partial T} \right)_K +
\left( {\partial L \over \partial K} \right)_T
\left( {\partial K \over \partial T} \right)_{m_q},
\label{6b}
\ee
must diverge at $T\!=\!T_{\x}$ in the chiral limit $m_q\!=\!0$, since the
chiral susceptibilities $\x_m^K = (\partial K / \partial m_q)_T$ and
$\x_T^K = (\partial K/\partial T)_{m_q}$ diverge in this limit.

\medskip

\begin{figure}[htp]
\centerline{\psfig{file=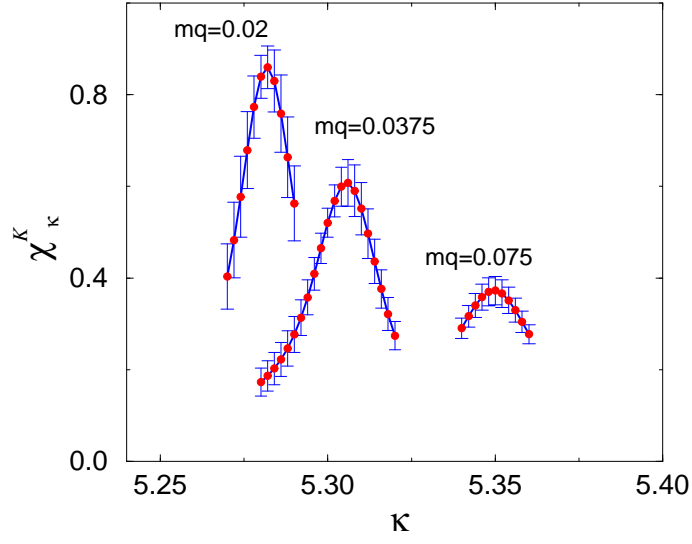,width=10cm}}
\vskip0.5cm
\caption{The chiral temperature susceptibility $\x^K_{\kappa}$ as function
of the temperature variable $\kappa=6/g^2$.}
\label{Dfig1}
\end{figure}

\begin{figure}[htp]
\centerline{\psfig{file=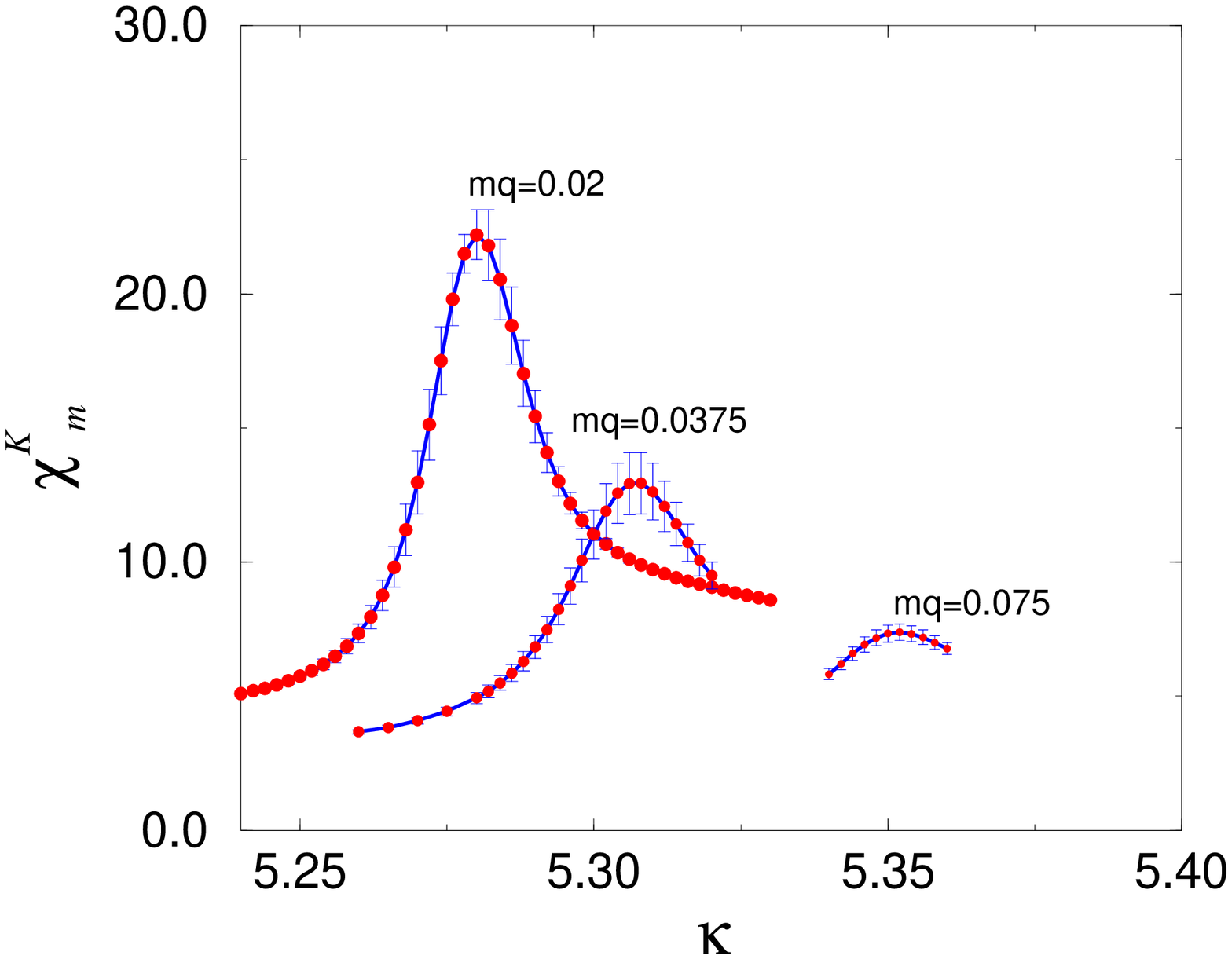,width=10cm}}
\vskip0.5cm
\caption{The chiral quark mass susceptibility $\x^K_m$ as
function of the temperature variable $\kappa=6/g^2$.}
\label{Dfig2}
\end{figure}

\begin{figure}[htb]
\centerline{\psfig{file=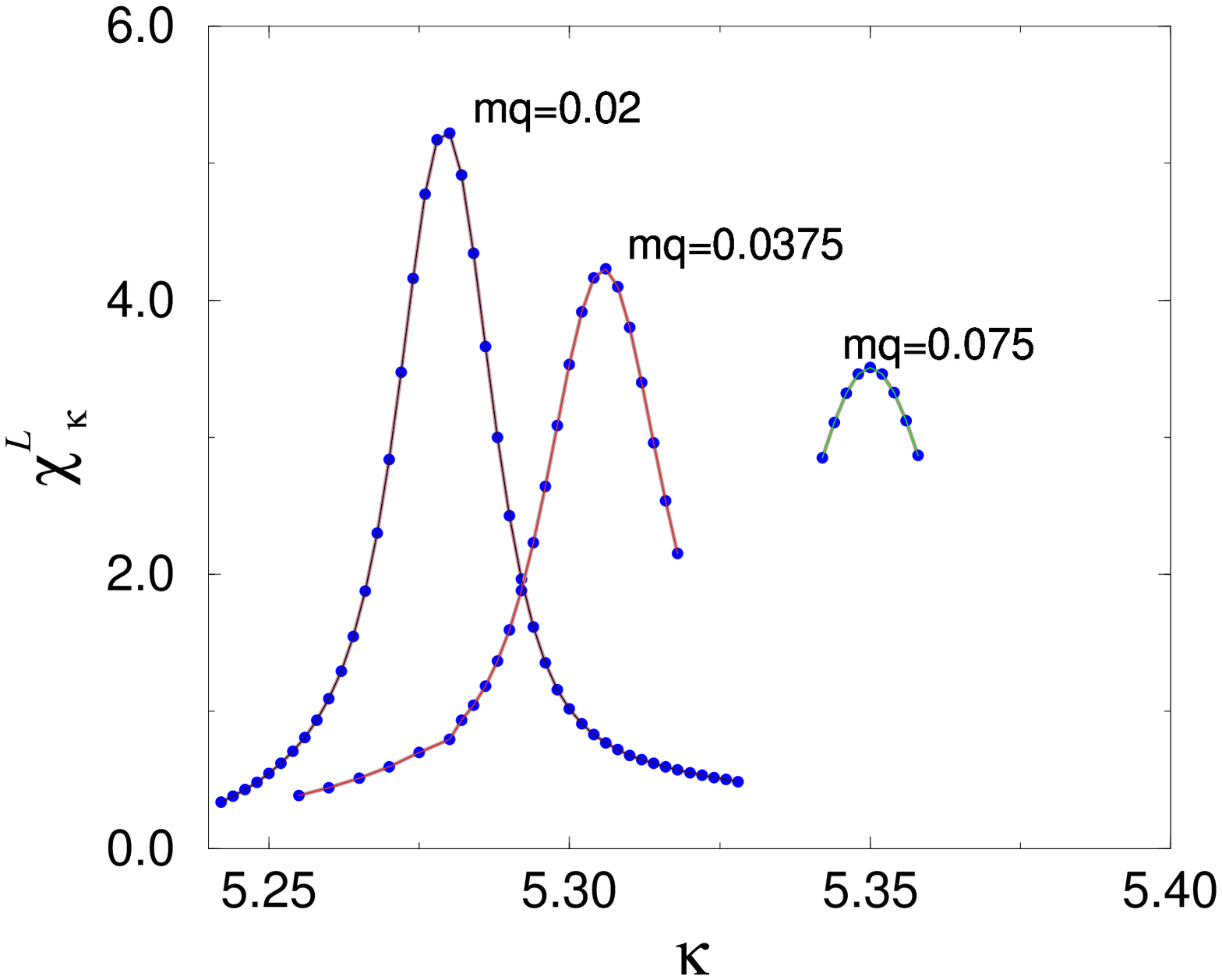,width=10cm}}
\vskip0.3cm
\caption{The Polyakov loop temperature susceptibility $\x^K_{\kappa}$ as
function of the temperature variable $\kappa=6/g^2$.}
\label{Dfig3}
\end{figure}

\begin{figure}[htb]
\centerline{\psfig{file=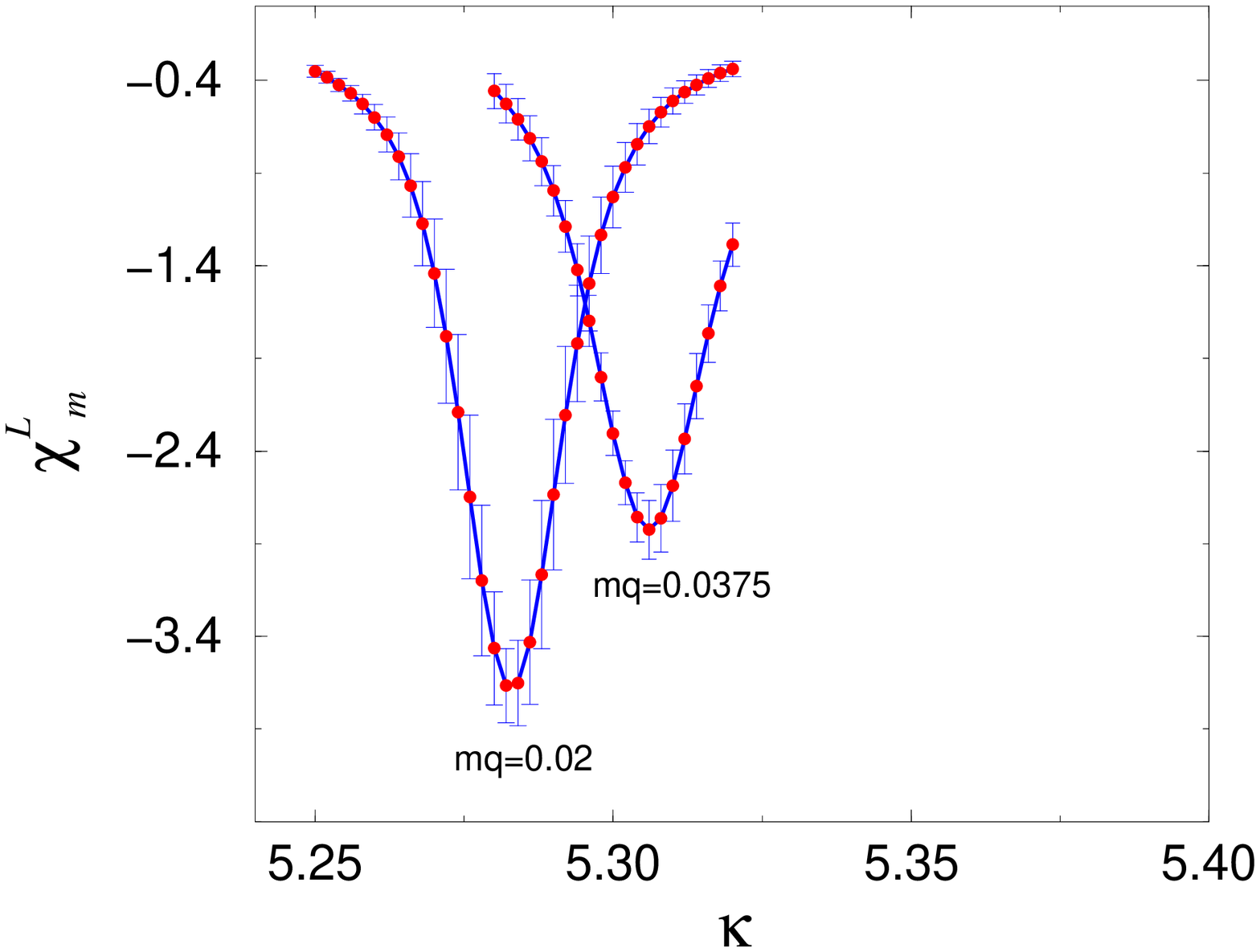,width=10cm}}
\vskip0.3cm
\caption{The Polyakov loop quark mass susceptibility $\x^K_m       $ as
function of the temperature variable $\kappa=6/g^2$.}
\label{Dfig4}
\end{figure}

In \cite{KL}, the chiral susceptibilities $\x_m^K$ and $\x_T^K$
were studied on an $8^3\times 4$ lattice for quark masses
$m_qa= 0.075,~0.0375$ and 0.02. In Figs.\ \ref{Dfig1} and \ref{Dfig2},
the results are shown as functions of the effective temperature
variable $\kappa=6/g^2$, where $g$ denotes the coupling in the QCD
Lagrangian. The increase of the peak height for decreasing quark mass
indicates the divergence in the chiral limit $m_q \to 0$. In Figs.
\ref{Dfig3} and \ref{Dfig4}, we show the corresponding results for the
Polyakov loop susceptibilities. They also peak sharply, the peak
positions coincide with those for the chiral susceptibilities, and here
as well the peak height increases with decreasing quark mass. The
observed behaviour therefore provides clear support for the divergence
of the temperature and quark mass derivatives of the Polyakov loop at
$T_{\x}$ in the chiral limit.

\medskip

A direct comparison of the functional behaviour, ideally of the relevant
critical exponents, for the two cases becomes difficult for two
reasons. The chiral susceptibilities do not at present lead to the
predicted $O(4)$ exponents; this may indicate that the quark masses are
still too large for pure critical behaviour. The Polyakov loop
susceptibilities contain in addition unknown non-singular factors
$(\partial L/ \partial T)_K$ and $(\partial L/ \partial K)_T$, which 
will modify the non-singular $m_q$-dependence relative to that of the 
chiral susceptibilities. Lattice studies for smaller quark masses would
therefore be very helpful.

\medskip

\begin{figure}[htb]
\centerline{\psfig{file=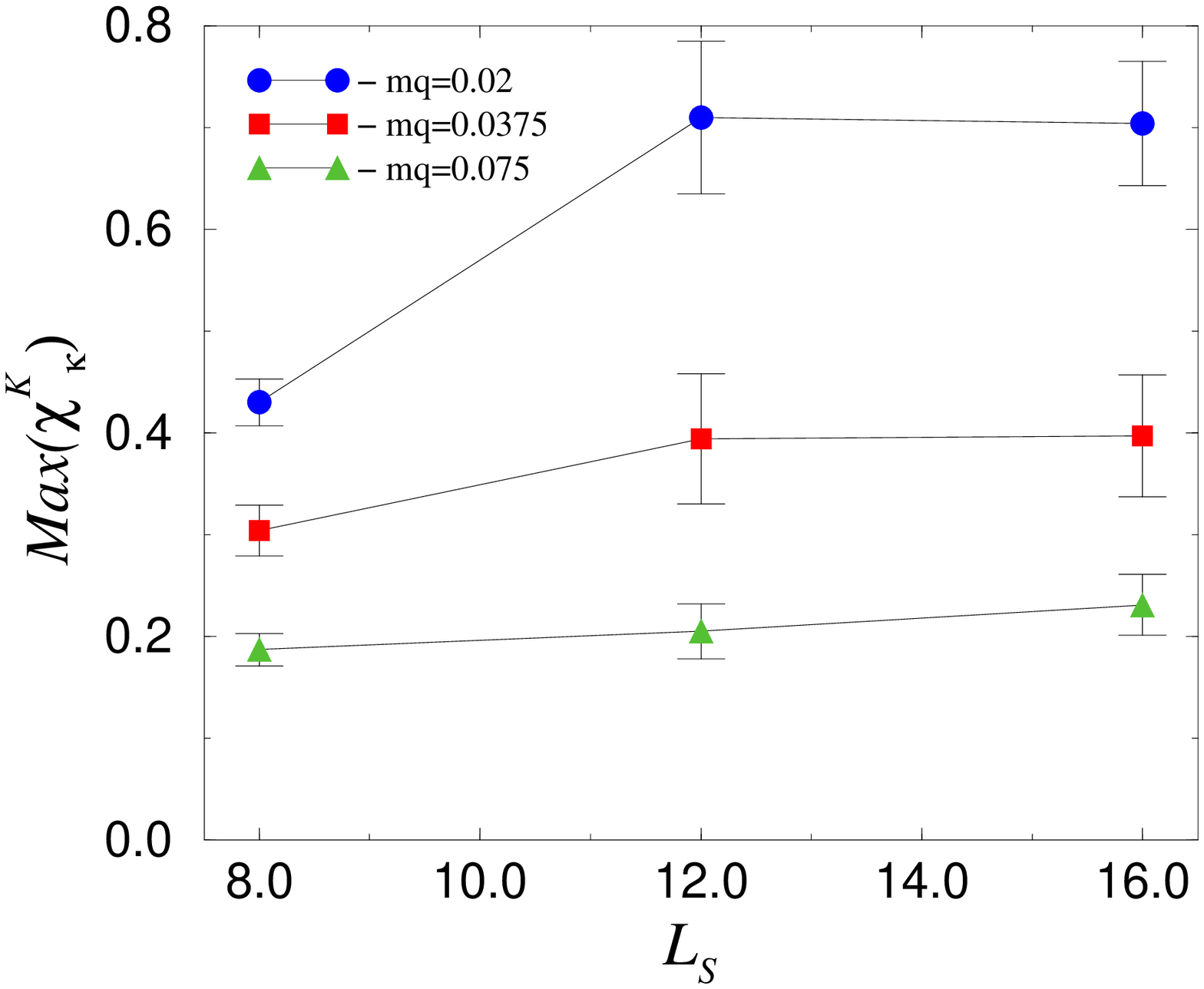,width=10cm}}
\vskip0.3cm
\caption{ The dependence of the $\x^K_{\kappa}$ peak height on spatial
lattice size.}
\label{Dfig5}
\end{figure}

\begin{figure}[htb]
\centerline{\psfig{file=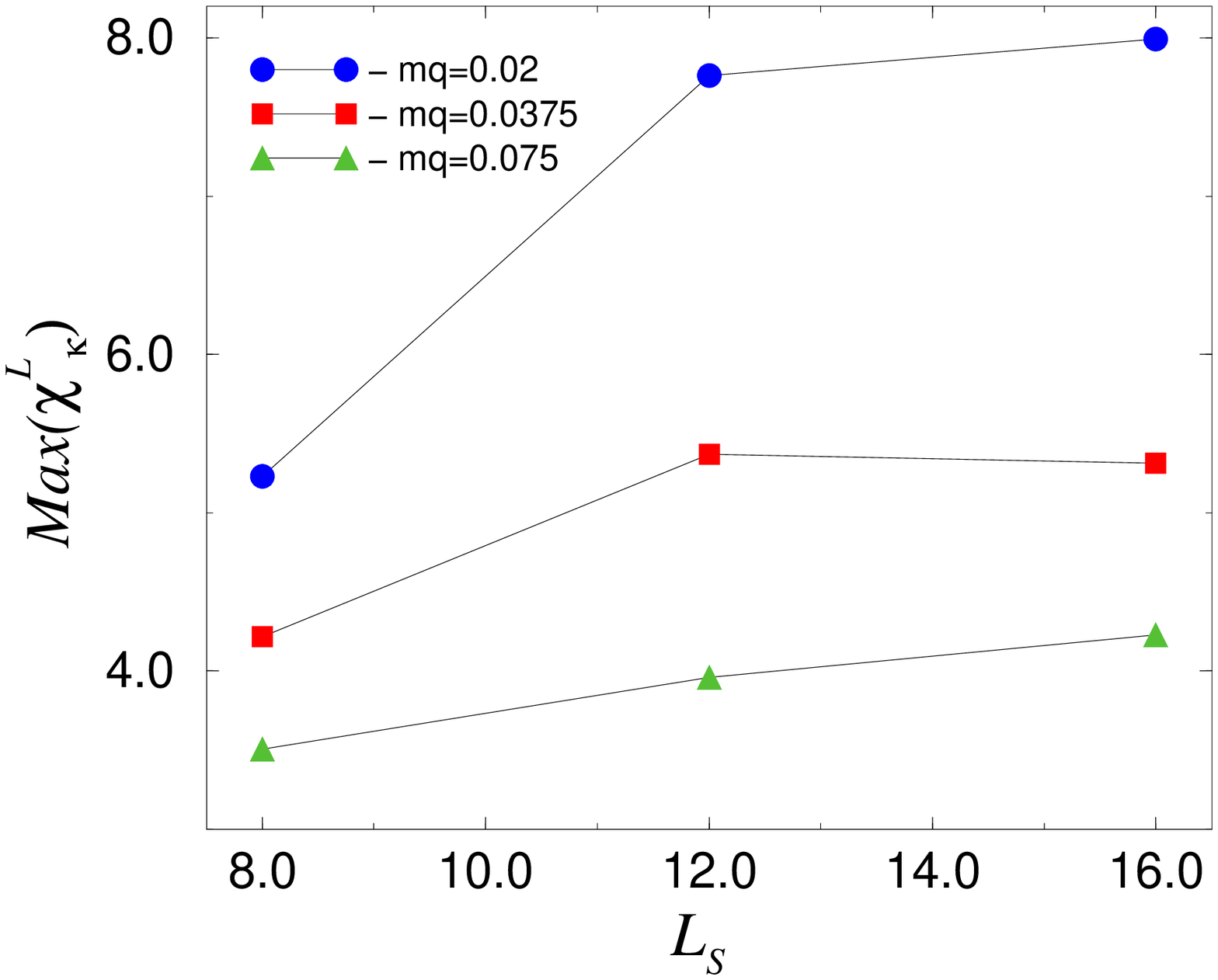,width=10cm}}
\vskip0.3cm
\caption{The dependence of the $\x^L_{\kappa}$ peak height on spatial
lattice size.}
\label{Dfig6}
\end{figure}

The `temperature' susceptibilities $\x^K_{\kappa}$ and $\x^L_{\kappa}$
have also been studied on larger lattices $12^3\times 4$ and $16^3
\times 4$, for the same three quark mass values \cite{EL}. In Figs.\
\ref{Dfig5} and \ref{Dfig6} we show the dependence of the peak height
on the spatial volume. The two susceptibilities show a very similar
peak increase in going from $8^3$ to $12^3$, but then saturation,
indicating that there are no further finite size effects for the given
quark mass value.

\medskip

In summary, we have shown that chiral symmetry restoration, with the
resulting sudden change in an effective constituent quark mass, leads
to a suddenly increasing external field which aligns the Polyakov loops
at $T=T_{\x} < T_d$
and thus produces a strong explicit breaking of the $Z_N$ symmetry of
the gauge field part of the Lagrangian. This effect makes chiral
symmetry restoration `coincide' with deconfinement. The resulting
predictions for diverging Polyakov loop susceptibilities are found to be
well supported by finite temperature lattice calculations for full QCD
with two flavours of light quarks.

\bigskip

\centerline{\bf Acknowledgements}

\bigskip

It is a pleasure to thank F.\ Karsch, P.\ Petreczky and A.\ M.\ Srivastava
for helpful discussions, and we are grateful for the financial support 
by the German Science Ministry BMBF under contract 06BI902.

~~~\vskip1cm

\end{document}